\documentstyle[preprint,aps,prb]{revtex}
\draft
\begin{document}
\bibliographystyle{plain}
\title{
Phase diagram of the one-dimensional extended
Hubbard model with attractive and/or repulsive interactions at quarter filling
}
\author {Karlo Penc$^{(*)}$}
\address{
     Institut de Physique, Universit\'e de Neuch\^atel\\
     1 Rue Breguet, CH-2000 Neuch\^atel (Switzerland)
}
\author {Fr\'ed\'eric Mila}
\address{
  Laboratoire de Physique Quantique\\
   Universit\'e Paul Sabatier, 31062 Toulouse, France
}

\maketitle

\begin{abstract}
We study the phase diagram of the one dimensional (1D) $U-V$ model at
quarter filling in the most general case where the on-site and first-neighbour
interactions $U$ and $V$ can be both attractive and repulsive. The results have
been obtained using
exact diagonalization of small clusters and variational techniques, as well as
exact results in various limits. We have analyzed four properties of the
groundstate: i)~whether it is insulating or metallic; \hbox{ii)~whether} it is
homogenous or phase separated; iii)~whether it has a spin gap; iv)~whether it
has dominant superconducting fluctuations. With eight phases, the resulting
phase diagram is unexpectedly rich. The four phases not found in the weak
coupling limit are: i) an insulating phase when $U$ and $V$ are large enough;
ii) a region of phase separation when $V$ is attractive; iii) another region of
phase separation when $V$ is large enough and $U$ small; iv) a region with
dominant superconducting fluctuations when $V$ is intermediate and $U$ small.
The actual nature of this
last phase, which has pairs but no spin gap, is not fully clear yet.
\end{abstract}

\newpage
\section{Introduction}

The phase diagram of one-dimensional models of correlated fermions is well
understood in the limit of weak interactions. In that case, the model can
be mapped using perturbation theory and a cut-off procedure onto the so-called
Fermi gas model (also referred to as $g$-ology), a model whose properties are
well established\cite{soly,emer}. In the opposite limit of strong interactions,
perturbation
theory cannot be used to perform such a mapping, and the determination of the
phase diagram is in most cases a very difficult problem. The exceptions are
models that have an exact solution, like for instance the 1D Hubbard
model\cite{lieb}.
In the case of that particular model, no qualitatively new physics appears in
going from weak to strong coupling.  So it is legitimate to wonder whether
there is a real need of studying the strong coupling regime of
more complicated models. We believe this is the case for at least two reasons.
First, real materials whose properties are of one-dimensional character are not
necessarily in the weak-coupling regime, and understanding their electronic
properties requires some knowledge of the properties of intermediate or strong
coupling models.
A very important example is the case of the organic conductors
of the (TMTSF)$_2$X family. In a large temperature range, they exhibit
low-energy properties that have a clear one-dimensional character and that are
compatible
with the Luttinger liquid picture\cite{wzie,dard,schu} if the exponent $\alpha$
defined by $N(\omega)-N(\omega_f) \propto \Theta(\omega_f-\omega)
(\omega-\omega_f)^\alpha$ is
slightly larger than $1$, where $N(\omega)$ is the density of states.
Such an exponent is impossible to get with weak interactions which
always yield $\alpha\ll 1$. The only hope to reach such a value is to turn to
strong coupling models. Let us note that the Hubbard model is not able either
to reproduce such an exponent: $\alpha$ is at most equal to $1/8$, a value
attained in the very strong coupling limit $U=+\infty$\cite{schu,paro}.

Second, it has been known for a long time that the strong coupling regime of
extensions of the Hubbard model can give rise to new physics. For example, it
was first shown by Ovchinnikov\cite{ovch} that adding a repulsion term $V$
between nearest neighbours to the Hubbard model yields a Metal-Insulator
transition for $V=2t$ when $U$ is infinite in the case of quarter-filling, a
phenomena not observed in the weak coupling regime of the same model. More
generally, commensurability between the number of particles and the number of
sites can give rise to a Metal-Insulator transition in the weak-coupling limit
only at half-filling, while this effect can in principle occur for any
commensurate filling if strong interactions are considered.

In this paper, we concentrate on the properties of the $U-V$ model at quarter
filling defined by the Hamiltonian
\begin{equation}
H=-t\sum_{i,\sigma}(c{}^\dagger_{i\sigma}c{}^{\ }_{i+1\sigma}+h.c.)
+U\sum_i n_{i\uparrow}n_{i\downarrow}
+V\sum_i n_in_{i+1}
\label{1}
\end{equation}
where $U$ and $V$ are allowed to take positive or negative values, i.e. to
describe repulsive or attractive interactions. An account of the properties of
this model in the case of repulsive interactions has already been
given\cite{mila}. Unexpected results showed up in the limit $U$ small, $V$
large, and an extension of the model to negative couplings was clearly needed
in order to clarify the properties of the model in the large $V$ limit.

Before we start the description of the phase diagram of the model of Equation
(\ref{1}), let us recall what is meant by phase diagram in the context of
one-dimensional systems\cite{soly,emer}. In general, phases are distinguished
by some kind of long-range order. If one sticks to that definition, there are
only three possibilities for 1D systems: i) Insulating phase if there is a gap
$\Delta_c$ in the charge sector; ii) Metallic phase if there is no gap in the
charge sector; iii) Phase separation if the ground-state is not homogenous. Any
kind of magnetic or superconducting long-range order is forbidden in 1D because
of quantum fluctuations. However, the correlation functions do not decay
exponentially but as power laws, that is
very slowly, and it is useful to distinguish regions in the metallic phase
according to the correlation function that decays most slowly. This is useful
because any infinitesimal coupling between the chains, that will be present in
real materials, will introduce long-range order associated with that
correlation function. According to  $g$-ology, there are at least four
different types of metallic phases that can be distinguished by the value of
the critical exponent $K_\rho$ (to be defined below), and by the presence or
absence of a gap $\Delta_s$ in the spin sector. Quite generally,
superconducting fluctuations dominate if $K_\rho>1$, while charge and spin
density fluctuations
dominate if $K_\rho<1$. Now, it there is no spin gap, triplet and singlet
pairing correlation functions on one hand, charge density and spin density
correlation functions on the other hand, have the same power law behaviour, and
one has to look at logarithmic corrections to see that triplet pairing and spin
density fluctuations dominate over singlet and charge density fluctuations
respectively. If there is a spin gap, triplet pairing and spin density
fluctuations decay exponentially, and singlet pairing and charge density will
be the dominant fluctuations. To summarize, one can expect to find six types of
phases:

1) Metallic with dominant spin density fluctuations if $\Delta_c=\Delta_s=0$
and $K_\rho<1$;

2) Metallic with dominant triplet pairing fluctuations if $\Delta_c=\Delta_s=0$
and $K_\rho>1$;

3) Metallic with dominant charge density fluctuations if $\Delta_c=0$,
$\Delta_s>0$ and $K_\rho<1$;

4) Metallic with dominant singlet pairing fluctuations if $\Delta_c=0$,
$\Delta_s>0$ and $K_\rho>1$;

5) Insulating if $\Delta_c > 0$;

6) Phase separation if the compressibility of the homogenous phase is negative.

The paper is organized as follows. In Section II, we summarize the results one
can get from $g$-ology in the weak coupling regime. In Section III, we present
the general phase diagram and provide detailed explanations on how the various
boundaries have been determined. In Section IV, we give some exact results that
can be obtained in the $V=+\infty$ case and that are useful to check the
numerical determination of the boundaries. Finally, a detailed discussion of
the large $V$, small $U$ case is given in Section V. This includes a derivation
of an effective Hamiltonian in that limit and some numerical results of flux
quantization used to analyze the presence of pairs in the ground-state.

\section{The weak-coupling phase diagram}

  When the interaction terms are small compared to the Fermi velocity, one can
 rewrite the Hamiltonian in momentum representation and map it onto the
$g$-ology model.
  For the Hamiltonian of the Eq.~(\ref{1}) at quarter filling the $g$
parameters are given by
\begin{eqnarray}
  g_1 &=& U+2V\cos 2k_F a = U \nonumber\\
  g_2 &=& U+2V \nonumber \\
  g_3 &=& 0 \nonumber\\
  g_{4\bot}&=&U+2V \nonumber\\
  g_{4\Vert}&=&2V
\end{eqnarray}
The coupling $g_3$ is equal to zero because we are away from half filling.

The schematic phase diagram for the quarter--filled model is presented in
Fig.~\ref{fig:phadiagwc}, where the different phases are numbered from (1) to
(4).
  If $U>0$ ($g_1>0$), we scale to the Tomonaga-Luttinger model. From the
scaling invariant  $2g_2-g_1$ the fixed--point value of the forward scattering
is  $g_2^*=U/2+2 V$.  For  $U>-V/4$ there are strong charge- or spin-density
wave fluctuations (region numbered by (1) in Fig.~\ref{fig:phadiagwc}) and for
$U<-V/4$  ($g_2^*<0$)   large superconducting fluctuations are present (phase
(2)).
On the other hand, if $U<0$ ($g_1<0$), we scale to strong coupling. We know
from the solution along the Luther-Emery line that there is a gap in the spin
spectrum, and charge-density (3) or superconducting fluctuations (4) are
favoured when $U+V/4$ is positive or negative respectively.

\section{The general phase diagram}

To determine the general phase diagram, we must calculate the charge gap
$\Delta_c$, the spin gap $\Delta_s$, the critical exponent $K_\rho$ and the
compressibility $\kappa$.

\subsection{The charge gap}

Let us start with the charge gap. It is given by
$\Delta_c=\lim_{L\rightarrow +\infty} \Delta(L;L/2)$, where
$\Delta(L;N)$ is defined by
\begin{equation}
\Delta(L;N)=E_0(L;N+1)+E_0(L;N-1)-2E_0(L;N)
\label{2}
\end{equation}
In that expression, $E_0(L;M)$ is the ground-state energy of $M$  particles on
$L$ sites and can be
obtained by exact diagonalization of small clusters using Lanczos algorithm.
Such a procedure has been used by Spronken et al\cite{spro} for the case of
spinless fermions with nearest neighbour repulsion,
and the critical value $V_c=2t$ could be reproduced quite accurately by fitting
$\Delta(L;N)$ with polynomials of $1/L$. In our case,
the Hilbert space is much larger, and we have results
for 3 sizes only ($L$=8,12,16). So, we cannot expect a very good accuracy. In
fact, we could not get meaningful results by fitting these 3 values with
$A+B/L+C/L^2$. However, fitting any pair of values with $A+B/L$ gives
reasonable results that do not depend too much on the pair chosen to do the
fit. Besides, as far as the large $U$ case is concerned, the best result is
obtained by fitting the results for $L=8$ and $16$.
This is probably due to the fact that the $12$ site system is quite different
from the other two: To get smooth results as a function of $L$, one must
choose the boundary conditions so that $k_F$ is one of the allowed $k$ values,
that is periodic ones for 8 and 16 sites, and antiperiodic
ones for 12 sites\cite{spro}.
It turns out that the insulating region is confined to the $U>0,V>0$ sector.
The resulting boundary goes from $(U,V)=(+\infty,2t)$ to $(U,V)=(4t,+\infty)$.
The first point is just the exact result obtained by Ovchinnikov some years
ago\cite{ovch}, while the second one is an exact result that we derive in
Section III.

\subsection{The spin gap}

We now turn to the spin gap. It corresponds to the energy needed to make a
triplet excitation from the singlet ground-state and can be calculated as
$\Delta_s=\lim_{L\rightarrow +\infty}\Delta_s(L;L/2)$, where $\Delta_s(L;N)$ is
defined by
\begin{equation}
\Delta_s(L;N)=E_0(L;N;S^z=1)-E_0(L;N;S^z=0)
\label{3}
\end{equation}
$E_0(L;N;S^z=s)$ is the ground state energy of a system of $N$ particles on $L$
sites in the sector $S^z=s$. As for the charge gap, and presumably for the same
reasons, the best way of extrapolating the finite-size results is to do a
linear fit of the results obtained for $L=8$ and $L=16$. The resulting boundary
goes through the point $(U,V)=(0,0)$ with a slope parallel to the $V$ axis, as
it should according to the weak coupling results. Besides, our numerical
results seem to indicate that it ends at the point $(U,V)=(-4t,+\infty)$, a
result we also prove in Section III.

\subsection{The critical exponents}

The nature of the ground-state and of the low-lying excitations when there is
no charge gap is a very difficult problem. In principle, one can get the
excitation spectrum for finite systems and see what the elementary excitations
are. Hopelessly, with the sizes that are available, it is not possible to
distinguish any structure that will remain in the thermodynamic limit with any
degree of confidence. The only thing that is possible at the moment is to
assume that the spectrum has a certain structure and to check if this is
compatible with the finite-size results. In the case of 1D systems, we know
that, in the weak coupling regime, the metallic phase is a Luttinger liquid.
Noting that this is also true for the Hubbard model for all $U$, it is a
reasonable hypothesis to assume that this is still the case for the model of
equation (\ref{1}). Then one can deduce the parameters of the model from the
low energy part of the spectrum\cite{hald,frah} obtained numerically for
small clusters, and check a posteriori if
these parameters are consistent with the hypothesis. One can estimate the ratio
$u_\rho/K_\rho$ by using its relation to the compressibility $\kappa$
\begin{equation}
{\pi \over 2} {u_\rho \over K_\rho} =
{1\over n^2\kappa}
\label{4 a}
\end{equation}
\begin{equation}
\kappa={L\over N^2} \biggl({E_0(L;N+2)+E_0(L;N-2)-2E_0(L;N) \over
4}\biggr)^{-1}
\label{4 b}
\end{equation}
Equation (\ref{4 b}) is the
finite-size approximation to the compressibility,
$E_0(L;N)$ being the ground-state energy calculated with suitable
boundary conditions. The
charge velocity can be obtained from the low-energy spectrum as
\begin{equation}
u_\rho=(E_1(L;L/2;S=0)-E_0(L;L/2))/(2\pi/L)
\label{5}
\end{equation}
where $E_1(L;L/2;S=0)$ is the energy of the first singlet excited state.
The analysis of the $t-J$ model by Ogata et al\cite{ogat} was based on these
equations.

In our study, we have also used another relation that holds for Luttinger
liquid, namely
\begin{equation}
\sigma_0=2u_\rho K_\rho
\label{6 a}
\end{equation}
where $\sigma_0$ is the weight of the Drude peak of the conductivity.
In 1D systems, $\sigma_0$ can be obtained simply from\cite{kohn,zoto,shas}
\begin{equation}
\sigma_0={\pi \over L} {\partial^2 E_0(\phi) \over
\partial \phi^2} \Biggr |_{\phi=0}
\label{6 b}
\end{equation}
\noindent where $E_0(\phi)$ is the ground-state energy as a function of
a phase $\phi$ due to a flux $\Phi=L\phi$ through the ring.
Equations (\ref{4 a} -\ref{6 b}) provide us with 3
independent conditions on $u_\rho$ and $K_\rho$. This is very useful for
several
reasons. First, the conductivity is much simpler to evaluate numerically than
the compressibility, which involves systems with $N+2$ particles.  Besides, we
have good reasons to believe that equation (\ref{4 b}) does not give a reliable
estimate of the compressibility when $V$ is large, and it is important to be
able to determine $K_\rho$ without its help.

It turns out that there are two disconnected curves $K_\rho=1$. One curve goes
through the point $(U,V)=(0,0)$ and corresponds to the boundary found in the
weak-coupling case. The slope found at $(U,V)=(0,0)$ in our numerical
determination of this boundary is in good agreement with the one predicted by
the weak-coupling analysis. There is however another line $K_\rho=1$ in the
small $U$, large $V$ region which has no counterpart in the weak coupling case.
This line goes through the point $(U,V)=(0,\sim 8t)$ and seems to extend toward
the points $(4t,+\infty)$ and $(-4t,+\infty)$. It is not possible to decide
whether these points really correspond to the limits of this line on the basis
of our numerical results however:
The estimate of $K_\rho$ becomes unreliable when $V$ gets very large.
As we couldn't get exact results about $K_\rho$ in the $V=+\infty$ case, this
behaviour remains a conjecture. Our numerical results also suggest that this
line lies entirely in the region where there is no spin gap, although we cannot
be totally sure either.

To complete the analysis of the metallic phase, we now have to check the
validity of the hypothesis we made about the excitations of the system, namely
that they are those of a Luttinger liquid. We can check this hypothesis in two
independent ways.
First, we can use the consistency of the three equations used to determine
$K_\rho$ and $u_\rho$ to
check the assumption that the system is a Luttinger liquid:
Equations (\ref{4 a}-\ref{4 b}) and (\ref{6 a}-\ref{6 b}) are typical of
Luttinger liquids and will be violated if we have an
instability to another phase (e.g. CDW insulator). A convenient way to measure
the consistency of the three conditions is to calculate the ratio
$\sigma_0 / \pi n^2 \kappa u^2_\rho$
which equals 1 for a Luttinger liquid. We have performed a systematic
evaluation of this ratio along lines in the ($U,V$) plane.
In the region where equation (\ref{5}) can be used, i.e. when $V$ is not too
large,
we found that, to a
good accuracy, this ratio equals 1 in the metallic phase, that it
drops rapidly when one enters the insulating phase, and that the transition
point was in good agreement with our previous determination of the phase
boundary.

Another way to check that we have a Luttinger liquid is to extract the central
charge from the finite-size scaling of the ground-state energy density
\cite{frah}
\begin{equation}
e_0(L)=e_0(+\infty)-{\pi(u_\rho + u_\sigma) \over 6 L^2} c +o(1/L^2)
\label{7}
\end{equation}
$u_\sigma$ was obtained from
\begin{equation}
u_\sigma=(E_1(L;L/2;S^z=1) -E_0(L;L/2))/(2\pi/L)
\label{7a}
\end{equation}
where $E_1(L;L/2;S^z=1)$ is the energy of the first excited state in the sector
$S^z=1$. Comparing our results for $L$=8, 12 and 16, we found that this scaling
form was accurate except when $V$ is too large, and that the
central charge $c$ determined in this way equals 1 within $2\%$, in good
agreement with the exact value $c=1$ for Luttinger liquids.

To summarize, we have been able to prove numerically that, when it is metallic,
 the system is a Luttinger liquid, except when $V$ is large, in which case
finite-size effects prevent us from checking this hypothesis.

\subsection{The compressibility}

Finally, we must calculate the compressibility to determine whether the
homogenous phases considered so far are stable against phase separation.
Reliable estimates of the compressibility are very difficult to extract in the
interesting regions where it changes signs due to huge finite size effects. So
we had to turn to variational techniques to estimate this quantity.
More precisely, we have used an extended Gutzwiller ansatz to calculate the
ground-state energy as a function of density. Details can be found in Appendix
A. As for $K_\rho$, we found two disconnected lines of phase separation. One is
entirely located in the $V<0$ half plane and goes from $(U,V)=(+\infty,-t)$ to
$(U,V)=(-\infty,0)$. The first   point is exactly the value obtained by other
authors from the mapping of that model onto the anisotropic XXZ Heisenberg
model in the limit $U=+\infty$ (see e.g. Ref.~\onlinecite{hald}). The other
limit is not surprising either: For
$U=-\infty$, the groundstate is a collection of static pairs. If $V>0$, they
will tend to stay apart, while if $V<0$, they will gain energy by staying as
close as possible. The other phase separation line lies in the small $U$, large
$V$ region. It goes through the point $(U,V)=(0,15t)$ and apparently extends
toward the points $(U,V)=(4t,+\infty)$ and $(U,V)=(-4t,+\infty)$. In fact, we
will prove in the next Section that these points lie on the boundary. So this
variational approach to the determination of phase separation looks quite
reliable.

\subsection{The phase diagram}

The resulting phase diagram is shown in Fig.~\ref{fig:phadiag}. It contains
eight different phases. The physical origins of most of them is already clear.
The insulating phase is a spin density wave that occurs because of
commensurability and was discussed in some details before\cite{mila}. The
nature of the
ground-state in the $V<0$ phase separation region is also quite easy to guess.
If $U$ is large enough, one will have a region of density 1 with one particle
per site and an empty region, while if $U$ is small or negative, one will have
a region of density 2 and an empty region. The limit between the two cases is
presumably the continuation of the $\Delta_s=0$ line, although we have not
tried to determine it. Besides, four of the metallic phases have a
weak-coupling counterpart, and
they simply correspond to cases 1) to 4) of the Introduction. The two phases
that occur when $V$ is large and $U$ is small are not so easy to account for
however. The rest of the paper is devoted to a detailed discussion of this
region. In particular, the origin of the phase separation will be clear from
the solution of the $V=+\infty$ case.

\section{The $V=+\infty$ case}

  In this limit a pair of electrons on the same site cannot be broken, since it
would mean hopping to the nearest site at an energy cost of $V=+\infty$. The
treatment of the model in this limit is greatly simplified by this fact. The
unpaired fermions can move in the regions located between pairs. Moreover,
unpaired fermions cannot be on neighbouring sites, so the spin is not relevant
and its contribution is only to increase the degeneracy of the energy levels.
In Appendix B an effective Hamiltonian is derived for the case where the $V$ is
much larger than $U$ and $t$, and the $V=+\infty$ case is described by the
$H_0$ (Eq.~\ref{eq:H0}).

The eigenstates can be classified according to the number of pairs $M$, to
their position $i_m$ ($1\le i_m \le L$, $1\le m \le M$) and to the number of
unpaired electrons $N_m$ between pairs $m$ and $m+1$. The number of sites
between pairs $m$ and $m+1$ is given by $L_m=i_{m+1}-i_m-1$. Now the essential
point is that the subsystem made of $N_m$ unpaired fermions located between
pairs $m$ and $m+1$ has the same energy levels as a system of $N_m$ free
spinless fermions on a finite chain of length $L_m-N_m-1$. This can be shown in
two steps: i) there is a one to one correspondence between the states, ii) the
matrix elements of the Hamiltonian are the same after mapping.

   The reduction of size is due to  the constraint imposed by $V=+\infty$,
which implies that two fermions cannot be on nearest neighbour sites and thus
that the number of available configurations is greatly reduced. Now, if we
consider the entity made of a fermion and the site right to it as a new
particle, the number of available sites is reduced by the number of fermions
between the two pairs and by 1 because of the pair at the boundary. For
example, if we consider a state where there are 7 sites and two fermions
between two pairs, there are 6 configurations and the correspondence reads
\begin{eqnarray}
  |de\uparrow e \downarrow ee ede\rangle &\rightarrow&
|dsseed\rangle'\nonumber\\
  |de\uparrow ee \downarrow e ede\rangle &\rightarrow&
|dsesed\rangle'\nonumber\\
  |de\uparrow eee \downarrow  ede\rangle &\rightarrow&
|dseesd\rangle'\nonumber\\
  |de e\uparrow e\downarrow e ede\rangle &\rightarrow& |dessed\rangle'
\nonumber\\
  |de e\uparrow ee\downarrow  ede\rangle &\rightarrow&
|desesd\rangle'\nonumber\\
  |de ee\uparrow e\downarrow  ede\rangle &\rightarrow& |deessd\rangle'
\nonumber
\end{eqnarray}
where $e$, $d$ and $s$ stand for empty, doubly occupied and single occupied
sites.
Here the number of sites available for the free fermions is 4 instead of 7.

Then it is easy to see that these new particles obey the Pauli principle, and
that the action of $H_0$ is to let these new particles hop with an amplitude
$-t$ whenever there is a free site next to it. For instance,
\begin{equation}
  H_0 |de\uparrow ee\downarrow e e de\rangle =-t\left(
  |dee\uparrow e\downarrow e e de\rangle  +
  |de\uparrow e\downarrow e e e de\rangle  +
  |de\uparrow eee\downarrow  e de\rangle\right)
\end{equation}
can be translated as
\begin{equation}
  H'_0 |dsesed\rangle' =-t(
  |dessed\rangle'  +
  |dsseed\rangle'  +
  |dseesd\rangle' )
\end{equation}
So the energy levels are those of $N_m$ spinless fermions on $L_m-N_m-1$ sites.

 Let us remember that if we have $\tilde N$ spinless fermions on $\tilde L$
sites with open boundary conditions, then the states can be enumerated by
quantum numbers $q=1\dots\tilde N$ with the energies
$-2 t \cos q\pi /(\tilde L+1)$, so that the ground state energy is given by
\begin{eqnarray}
  E_{\rm sf}(\tilde L,\tilde N) &=&
     -2t \sum_{q=1}^{\tilde N}\cos {\pi q \over \tilde L+1} \\
 &=& t-t
  \left(\sin {\tilde N+1/2 \over \tilde L+1}\pi\right)
  \left(\sin {1 \over 2} {1 \over \tilde L+1}\pi\right)^{-1}
\end{eqnarray}

 Thus, the lowest energy of the configuration above is given as
\begin{equation}
  E = M U + \sum_i^M E_{\rm sf}(L_i-N_i-1,N_i)
\end{equation}

  Now we need to get the configuration of doubly occupied sites for which the
energy derived above is the lowest.
  One can easily check that if we divide a system of spinless fermions into
subsystems, so that the numbers of sites and particles are conserved, the
energy will always be higher than that of the original system. Note that this
result holds because we are considering open boundary conditions. As a
consequence the lowest energy is reached if all the pairs sit together. Hence,
the lowest energy for a system having $M$ pairs is given by
\begin{equation}
  E = M U - E_{\rm sf}(L-N,N-2M)
  \label{eq:ElU}
\end{equation}
where we have taken into account the fact that the effective space for the
$N-2M$ unpaired fermions is reduced from
$L-2M +1$ to $L-2M+1-(N-2M)-1=L-N$.

In the thermodynamic limit ($L\rightarrow\infty$) this is given by
\begin{equation}
  E/L = {m n \over 2} U
        - { 2 \over \pi}(1-n)
          \sin \pi {n\over 1-n} (1-m)
\end{equation}
where  $m=2M/N$ is the proportion of fermions forming pairs and $n=N/L$ is the
density of the fermions.

 This energy should be minimized as a function of $m$.  If $U>U_c=-4 t\cos \pi
n/(1-n)$
the energy is minimized for $m=0$ and there are no pairs in the system. If
$-4t<U<U_c$, the minimum is reached for $m$ given by
\begin{equation}
  m=  {2n-1\over n}+{1-n\over n}{1\over \pi}\arccos(-U/4t)
\end{equation}
Finally, when $U<-4t$ the energy is minimal for $m=1$, which means that all the
fermions are paired.

For $n=1/2$ (quarter filled system) $U_c=4t$. Above $4t$ every second site is
occupied by one fermion with spin up or down. This would correspond to a filled
band in our effective model of spinless fermions. However, if we create a pair,
it would cost us an energy $U$, and we can gain $-4t$ in kinetic energy, since
we have created two holes at the top of the filled band each having energy
$-2t$ (see Fig.~\ref{fig:vinf}).
Thus the creation of pairs becomes favorable when $U<4t$.
This estimate is in complete agreement with the value of $U_c$ obtained from
Eq.~\ref{2} for the charge gap, where $E_0(L,N)=0$ (Fig.~\ref{fig:vinf}a),
$E_0(L,N+1)=U$ (Fig.~\ref{fig:vinf}b) and $E_0(L,N-1)=-4t$
(Fig.~\ref{fig:vinf}c,d) and $\Delta_c=U-4t$ becomes zero when $U=4t$. This
proves that $(U,V)=(4t,+\infty)$  is on the boundary of the metal--insulator
transition.

  Let us now look at the spin gap. When $U<-4t$ we have only local spins in the
ground state, and $\Delta_s>0$ . When $U$ becomes larger than $-4t$, paired
fermions appear in the ground state. But we saw above that the energy is
independent of the spin, which means that $\Delta_s=0$. So
$(U,V)=(-4t,+\infty)$ lies on the boundary of the spin gap.

  Finally, it is clear from Eq.~(\ref{eq:ElU}) that there is phase separation:
The ground state consists of a domain of density 1 where all the pairs are
packed and a domain of density $<1/2$ where the unpaired particles can move.
So the points $(U,V)=(-4t,+\infty)$ and $(4t,+\infty)$ lie on the boundary of
the phase separation.

\section{Discussion of the large $V$ region}

The nature of phase (8) is quite puzzling. On one hand, our numerical results
indicate that this phase has no charge gap, no spin gap, and that $K_\rho$ is
larger than 1. Then, according to $g$-ology, the triplet pairing fluctuations
should dominate. On the other hand, if we try to understand why this phase
might be superconducting, our results concerning the $V=+\infty$ suggest that
it is due to the presence of local pairs in the ground-state. The argument goes
as follows: When $V=+\infty$ and $-4t<U<4t$, the ground-state consists of local
pairs that sit together, and of unpaired particles that can move freely (apart
from the $V$ interaction) in the part of the sample not occupied by pairs.
Then, when $V$ decreases, the pairs start moving around.
This can be seen by deriving an effective Hamiltonian in the limit of large but
finite $V$ (see Appendix B). The system clearly consists of two types of
carriers, local pairs and unpaired particles, that can hop, cross, or exchange.
We have not attempted yet a careful study of this effective Hamiltonian. One
scenario as to what happens when $V$ decreases is the following.
As long as $V$ is not too small, the pressure exerted by the free carriers is
sufficient to force the pairs to stay together, and there is still phase
separation. However, when $V$ is small enough, the kinetic energy gained by the
pairs is comparable to that of the free carriers, and they can move freely in
the sample. The ground-state is then homogenous and metallic. Superconductivity
would then arise from the presence of locals pairs. But these pairs are located
on a single site and must be singlet. Then this picture would favour singlet
superconductivity, in contradiction to the prediction of $g$-ology.

Let us note that there is no inconsistency so far. On one hand, we have not
been able to check that the system is a Luttinger liquid when $V$ is large, so
that the predictions of $g$-ology shouldn't be taken too seriously in that
region. On the other hand, we have no precise indication that there are still
pairs in the ground-state when $V$ is so small that we are outside the region
of phase separation. In fact, an alternative scenario to that outlined above
for the large $V$ case is that the
pairs disappear when one leaves the region of phase separation. Instead of
having to reconcile these points of view, it looks more like we have to choose
between one of the following possibilities: i) the system is a Luttinger
liquid, there is no pair in the ground-state, and the triplet pairing
fluctuations dominate; ii) the system has two types of carriers, local pairs
and unpaired particles, it is not a Luttinger liquid, and the singlet pairing
fluctuations dominate. How can we choose?

The most natural thing to do would be to look at the low-lying excitations of
the system to see whether there are of the Luttinger-liquid type or not. We
have indeed tried that, but with no success. For systems as small as 16 sites,
it is impossible to say anything by looking at the spectrum directly. There is
however an indirect way of testing whether there are pairs in the ground-state
by studying the response of its ground-state energy to a magnetic flux.

To be more specific, suppose that a closed ring is threaded by a magnetic flux
$\Phi$. The units are such that $\Phi=2\pi$ corresponds to one flux quantum
$hc/e$. This can be taken into account by replacing the kinetic energy by
\begin{equation}
 -t\sum_{i,\sigma}({\rm{e}^{i\Phi/L} }
  c^\dagger_{i\sigma}c^{\phantom{\dagger} }_{i+1\sigma}+h.c.)
\end{equation}
where $L$ is the number of sites.
Then, for a noninteracting system the function $f(\Phi)$ defined by
$f(\Phi)=\lim_{L\rightarrow\infty} L(E_0(\Phi)-E_0(0))$, where $E_0(\Phi)$ is
the ground state energy for a given flux, is periodic with period $2\pi$.
However, if the ground state consists of pairs, like for the
negative $U$ Hubbard
model, the function $f(\Phi)$ is periodic with period $\pi$. If the ground
state has both types of carriers, then we expect to be in the intermediate
situation where this function is periodic with period $2\pi$ but has a local
minimum at $\pi$.

 For finite systems, one finds almost systematically two minima. This is even
true in the case of the Hubbard model with repulsive $U$. The only meaningful
information is contained in the limiting behaviour when the system-size becomes
infinite. With systems consisting of 8, 12 and 16 sites, all we can see is the
trend. Some results are given in Fig.~\ref{fig:flux}. Quite clearly, when $U$
is negative, but still inside phase (8), the local minimum at $\pi$ gets
deeper when $L$ increases, which suggests that there are pairs in the
ground-state in the thermodynamic limit. For larger values of $U$, there is
still a well defined local minimum, but it tends to decrease with the size of
the system, so it is impossible to know whether it will survive in the
thermodynamic limit.

To summarize what we have learned about phase (8), it seems to us likely that
when $U<0$ there are pairs in the ground-state, that the system is not a
Luttinger liquid but has a more complicated spectrum including excitations of
both the pairs and the unpaired particles, and that it has strong singlet
pairing fluctuations. When $U$ becomes positive, the nature of this part of the
phase diagram is still very much an open question. There might still be pairs
around, so that the whole region has the same behaviour, but the pairs might
have disappeared as well, in which case the system would be a Luttinger liquid
with triplet pairing excitations. More work is needed to settle this issue.

\acknowledgements

We acknowledge useful discussions with H. Beck, P. Fazekas, T. Giamarchi, D.
Poilblanc, H. Schulz and X. Zotos.

This work was done while both authors were postdocs at the University of
 Neuchatel and
was supported in part by the Swiss National Science Foundation under Grants
Nos. 20-31125.91 and 20-33964.92.

\appendix

\section{Cluster Gutzwiller Approximation}

  The ground-state energy has been estimated with the help
of a variational wave function, namely a  Gutzwiller-like wave function
\cite{gutzw} extended by intersite correlations. That intersite correlations
might be essential was first pointed out by Kaplan {\it al.}\cite{kaplan}, who
showed that the correct $-t^2/U$ energy can be obtained in the case of the
half-filled Hubbard model by including the intersite empty-doubly occupied
correlation in addition to the on-site correlation. This kind of variational
wave function has been applied to different models, and it proved to be useful
to get results in regions where other methods have difficulties.

  The variational wave function we used can be written
\begin{equation}
  |\Psi_{EG}\rangle = \hat P |FS \rangle
\end{equation}
where the operator $\hat P$ contains the intersite correlations and is given by
\begin{equation}
\hat P=\prod_i\prod_{\mu,\nu}
   \bigl[
      1-(1-\lambda_{\mu\nu})\hat P_{\mu,i}\hat P_{\nu,i+1}
   \bigr]
\end{equation}
The projectors $\hat P_{\mu,i}$ at site $i$ are defined by
\begin{equation}
\hat P_{\mu,i}=
   \left\{
    \begin{array}{cl}
         (1-n_{\uparrow,i})(1-n_{\downarrow,i}),& \mbox{\quad if $\mu=e$;} \\
         n_{\uparrow,i}(1-n_{\downarrow,i}),& \mbox{\quad if $\mu=\uparrow$;}
\\
         (1-n_{\uparrow,i})n_{\downarrow,i},& \mbox{\quad if $\mu=\downarrow$;}
\\
         n_{\uparrow,i}n_{\downarrow,i},& \mbox{\quad if $\mu=d$.}
    \end{array}
    \right.
\end{equation}
The $\lambda_{\mu\nu}$'s are the weighting amplitudes of the different nearest
neighbour intersite correlations. For instance, if an empty site and a doubly
occupied site are nearest neighbours, the weight of the configuration is
multiplied by $\lambda_{ed}$. Altogether there are 16 different
$\lambda_{\mu\nu}$'s, but they are not independent. Different symmetries, like
$\lambda_{\mu\nu}=\lambda_{\nu\mu}$ and zero magnetization, reduce the number
of independent weighting amplitudes to 7. Furthermore, the density of particles
being given, there are constraints to be satisfied in the minimization process.

The difficult step is to calculate $\langle \Psi_{EG}| H|\Psi_{EG}\rangle$.
In general, this cannot be done analytically\cite{metzner}, and the only
"exact" way is to do it with variational Monte Carlo\cite{VMC}. There is
however an approximate way of calculating this matrix element known as the
cluster Gutzwiller Approximation\cite{cga} which leads to relatively modest
numerical tasks, and which has been shown to give very good results on various
models. In this scheme,
a cluster of four consecutive lattice sites is chosen, and they are labeled by
1,2,3 and 4. On this small cluster we can have $4^4=256$ different
configurations. The weight of each configuration is its weight in the Fermi sea
multiplied by the factors $\lambda$, and the matrix elements are  the matrix
elements taken in the Fermi sea multiplied by the amplitudes $\lambda$. For
instance,
\begin{eqnarray}
  \langle ed\uparrow\downarrow|ed\uparrow\downarrow \rangle
  & \rightarrow &
  \lambda_{ed}^2 \lambda_{d\uparrow}^2 \lambda_{\uparrow\downarrow}^2
  \langle ed\uparrow\downarrow|ed\uparrow\downarrow \rangle_{FS} \nonumber\\
 \langle ed\uparrow\downarrow|\hat A|ed\uparrow\downarrow \rangle
  & \rightarrow &
  \lambda_{ed}^2 \lambda_{d\uparrow}^2 \lambda_{\uparrow\downarrow}^2
  \langle ed\uparrow\downarrow|\hat A|ed\uparrow\downarrow \rangle_{FS}
\nonumber\\
 \langle ed\uparrow\downarrow|\hat A|e\uparrow d\downarrow \rangle
  & \rightarrow &
  \lambda_{ed}\lambda_{e\uparrow} \lambda_{d\uparrow}^2
\lambda_{\uparrow\downarrow}^2
  \langle ed\uparrow\downarrow|\hat A|ed\uparrow\downarrow \rangle_{FS}
\end{eqnarray}
The energy can be obtained by summing the matrix elements of the Hamiltonian
over all 256 configurations divided by the norm of the states.
 However, since the central pairs (2 and 3) and the shell pairs (1 and 2 , 3
and 4) are nor treated in a symmetric way, the cluster is not homogeneous. In
order to force homogeneity, different amplitudes $\lambda_{\mu\nu,s}$ and
$\lambda_{\mu\nu,c}$ can be defined for the shell and central pairs.
Then equation (A4) has to be modified according to
\begin{eqnarray}
  \langle ed\uparrow\downarrow|ed\uparrow\downarrow \rangle
  & \rightarrow &
  \lambda_{ed,s}^2 \lambda_{d\uparrow,c}^2 \lambda_{\uparrow\downarrow,s}^2
  \langle ed\uparrow\downarrow|ed\uparrow\downarrow \rangle_{FS} \nonumber\\
 \langle ed\uparrow\downarrow|\hat A|ed\uparrow\downarrow \rangle
  & \rightarrow &
  \lambda_{ed,s}^2 \lambda_{d\uparrow,c}^2 \lambda_{\uparrow\downarrow,s}^2
  \langle ed\uparrow\downarrow|\hat A|ed\uparrow\downarrow \rangle_{FS}
\nonumber\\
 \langle ed\uparrow\downarrow|\hat A|e\uparrow d\downarrow \rangle
  & \rightarrow &
  \lambda_{ed,s}\lambda_{e\uparrow,s} \lambda_{d\uparrow,c}^2
\lambda_{\uparrow\downarrow,s}^2
  \langle ed\uparrow\downarrow|\hat A|ed\uparrow\downarrow \rangle_{FS}
\end{eqnarray}
and in the minimization procedure we impose the constraint that the densities
and correlations are the same everywhere in the cluster. Details can be found
in Ref. \cite{cga}.

  Once we have the energy, the inverse compressibility $\kappa ^{-1}$ is easily
calculated as
\begin{equation}
  \kappa^{-1}={\partial^2 E \over \partial^2 n}
\end{equation}
We define the point of phase separation to be the point where $\kappa^{-1}$
vanishes, i.e. where the system becomes unstable against local density
fluctuations. A different definition of the phase separation boundary can be
given by Maxwell construction.

\section{Effective Hamiltonian for large $V$}
In this appendix, we derive an effective Hamiltonian in the limit where
the intersite repulsion $V$ is much larger than the on-site interaction $U$ and
the hopping $t$. This can be done by using the Schrieffer-Wolff
transformation\cite{schrieffer}
\begin{eqnarray}
  H_{\rm eff} &=& {\rm e}^{iS} H {\rm e}^{-iS} \nonumber\\
              &=& H + i \left[S,H \right] -
                   {1 \over 2} \left[S,\left[S,H \right]\right]+\dots
\end{eqnarray}
We first split the Hamiltonian into two parts,
$H=H'+H_{\rm mix}$, where $H'$ mixes states within a given "Hubbard Band",
while $H_{\rm mix}$ has matrix elements between different Hubbard bands
typically separated by an energy of order $U$ from each other.
 We can eliminate $H_{mix}$ by choosing $S$ so
that
\begin{equation}
H_{\rm mix}=-i \left[S,H' \right]
  \label{eq:SWcond}
\end{equation}
Then to first order in $S$
\begin{equation}
  H_{\rm eff} = H' + i {1\over 2} \left[S,H_{\rm mix} \right].
\end{equation}

Using standard procedures, the resulting effective Hamiltonian we obtained
consists of six parts
\begin{equation}
H_{\rm eff} = H_0+H_1+H_2+H_3+H_4+H_5
\end{equation}
The leading term $H_0={\cal P} H' {\cal P}$ reads
\begin{equation}
H_0 = {\cal P}
 \left[ -t
  \sum_{i,\sigma}
    \left(
      c^\dagger_{i+1,\sigma}c^{\phantom{\dagger}}_{i,\sigma}+{\rm h.c.}
    \right)
  + U \sum_{i} n_{i,\sigma} n_{i,-\sigma}
 \right]
 {\cal P}
 \label{eq:H0}
\end{equation}
where
\begin{equation}
  {\cal P} = \prod_{i,\sigma,\sigma'} (1-n_{i,\sigma} n_{i+1,\sigma'})
   \label{eq:constraint}
\end{equation}
This is a Hubbard model with the additional constraint that two fermions cannot
be on neighbouring sites.

The remaining $H_i$'s are of order $-t^2/V$.  $H_1$ describes
the hopping of pairs
\begin{equation}
  H_1= -2 {t^2 \over V} \sum_{i} d^\dagger_i d^{\phantom{\dagger}}_i +{\rm h.c}
\end{equation}
where $d^\dagger_i=c^\dagger_{i,\uparrow}c^\dagger_{i,\downarrow}$
($d_i=c^{\phantom{\dagger}}_{i,\downarrow}c^{\phantom{\dagger}}_{i,\uparrow}$ )
is the pair creation (annihilation) operator.
Annihilation (creation) of pairs into (from) opposite spin fermions two sites
apart is described by
\begin{equation}
  H_2= - {t^2 \over V} \sum_{i}
    \left(
       c^\dagger_{i+1,\uparrow}c^\dagger_{i-1,\downarrow}
      +c^\dagger_{i-1,\uparrow}c^\dagger_{i+1,\downarrow}
    \right)
    \left(
         d^{\phantom{\dagger}}_{i+1}
      +2 d^{\phantom{\dagger}}_{i}
      +  d^{\phantom{\dagger}}_{i-1}
    \right)
    + {\rm h.c.}
\end{equation}
$H_3$ describes a pair-electron interchange
\begin{equation}
  H_3 = {t^2 \over 2V}
    \sum_{i,\sigma}
        c^\dagger_{i-1,\sigma}
        d^\dagger_{i+1}
        d^{\phantom{\dagger}}_{i-1}
        c^{\phantom{\dagger}}_{i+1,\sigma}     + {\rm h.c.}
\end{equation}
while $H_4$ stands for a simultaneous hopping of two fermions
\begin{equation}
  H_4= - {t^2 \over V}
    \sum_{i}
        c^\dagger_{i+2,\sigma}
        c^{\phantom{\dagger}}_{i,\sigma}
        c^\dagger_{i+1,-\sigma}
        c^{\phantom{\dagger}}_{i-1,-\sigma}
       + {\rm h.c.}
\end{equation}
Finally, we have a diagonal term
\begin{equation}
  H_5= -{t^2 \over 6V}
  \sum_{i}
    \left[
      \left(
        n_{i-1,\uparrow}+n_{i-1,\downarrow}
       +n_{i+1,\uparrow}+n_{i+1,\downarrow}-4
      \right)^2+8
    \right]
\end{equation}

\begin{figure}
\caption{ Phase diagram in the weak coupling limit. \label{fig:phadiagwc}}
\end{figure}

\begin{figure}
\caption{ Phase diagram of the $U-V$ model at quarter-filling. The nature of
the various phases is discussed in the main text.
\label{fig:phadiag}}
\end{figure}

\begin{figure}
\caption{ Typical configuration for $N$ particles on $L$ sites in the
 $V=+\infty$ limit:
  a) $N=L/2$; b) $N=L/2+1$; c) and d) $N=L/2-1$. Dots stand for empty sites.
  \label{fig:vinf}}
\end{figure}

\begin{figure}
\caption{ Ground state energy as a function of flux for two cases:
  a) $U=-1$ and $V=8$ and  b) $U=0$ and $V=12$. The boundary conditions are
  periodic for $L=12$ and antiperiodic for $L=8$ and $L=16$. The curves having
a
  minimum at $\phi=0$ are nearly indistinguishble for $L=8,12$ and $16$, and we
  have plotted a single set of points.
\label{fig:flux}}
\end{figure}

\end{document}